\documentclass[sigconf,natbib=false]{acmart}
\usepackage{subcaption}
\DeclareUnicodeCharacter{2005}{\hspace{0.25em}}

\AtBeginDocument{%
  }

\acmConference[]{}{}{}
\acmDOI{}
\acmISBN{}
\acmYear{}

\RequirePackage[
  backend=biber,
  datamodel=acmdatamodel,
  style=acmnumeric,
  sorting=none,
  ]{biblatex}
\setcopyright{none}

\begin{document}

\title{Mycorrhizal Fungi and Plant Symbiosis for Energy Harvesting in the Internet of Plants}

\author{Fatih E. Bilgen\textsuperscript{*} \qquad Ozgur B. Akan\textsuperscript{*}\textsuperscript{\dag}}

\affiliation{%
  \institution{\textsuperscript{*}Ko\c c University Center for neXt-generation Communications~(CXC)}
  \department{Department of Electrical and Electronics Engineering}
  \institution{Ko\c c University, 34450, Istanbul}
  \country{T\"urkiye}
}
\email{{fbilgen20, akan}@ku.edu.tr}

\affiliation{%
  \institution{\textsuperscript{\dag}Internet of Everything (IoE) Group}
  \department{Electrical Engineering Division, Department of Engineering}
  \institution{University of Cambridge, CB3 0FA, Cambridge}
  \country{United Kingdom}
}
\email{oba21@cam.ac.uk}

\renewcommand{\shortauthors}{Bilgen et al.}

\begin{abstract}
Biological entities in nature have developed sophisticated communication methods over millennia to facilitate cooperation. Among these entities, plants are some of the most intricate communicators. They interact with each other through various communication modalities, creating networks that enable the exchange of information and nutrients. In this paper, we explore this collective behavior and its components. We then introduce the concept of agent plants, outlining their architecture and detailing the tasks of each unit. Additionally, we investigate the mycorrhizal fungi-plant symbiosis to extract glucose for energy harvesting. We propose an architecture that converts the chemical energy stored in these glucose molecules into electrical energy. We conduct comprehensive analyses of the proposed architecture to validate its effectiveness.
\end{abstract}


\maketitle

\section{Introduction}
Natural systems have evolved over millennia, undergoing significant refinement through extensive adaptation to diverse and challenging environments. A vital component of these systems is the plant kingdom. Plants play a pivotal role in maintaining ecological balance and are indispensable for life on Earth. They perform numerous functions that sustain the conditions necessary for the survival and prosperity of all living organisms.

\begin{figure}[t]
    \centering
    \includegraphics[scale = 0.2]{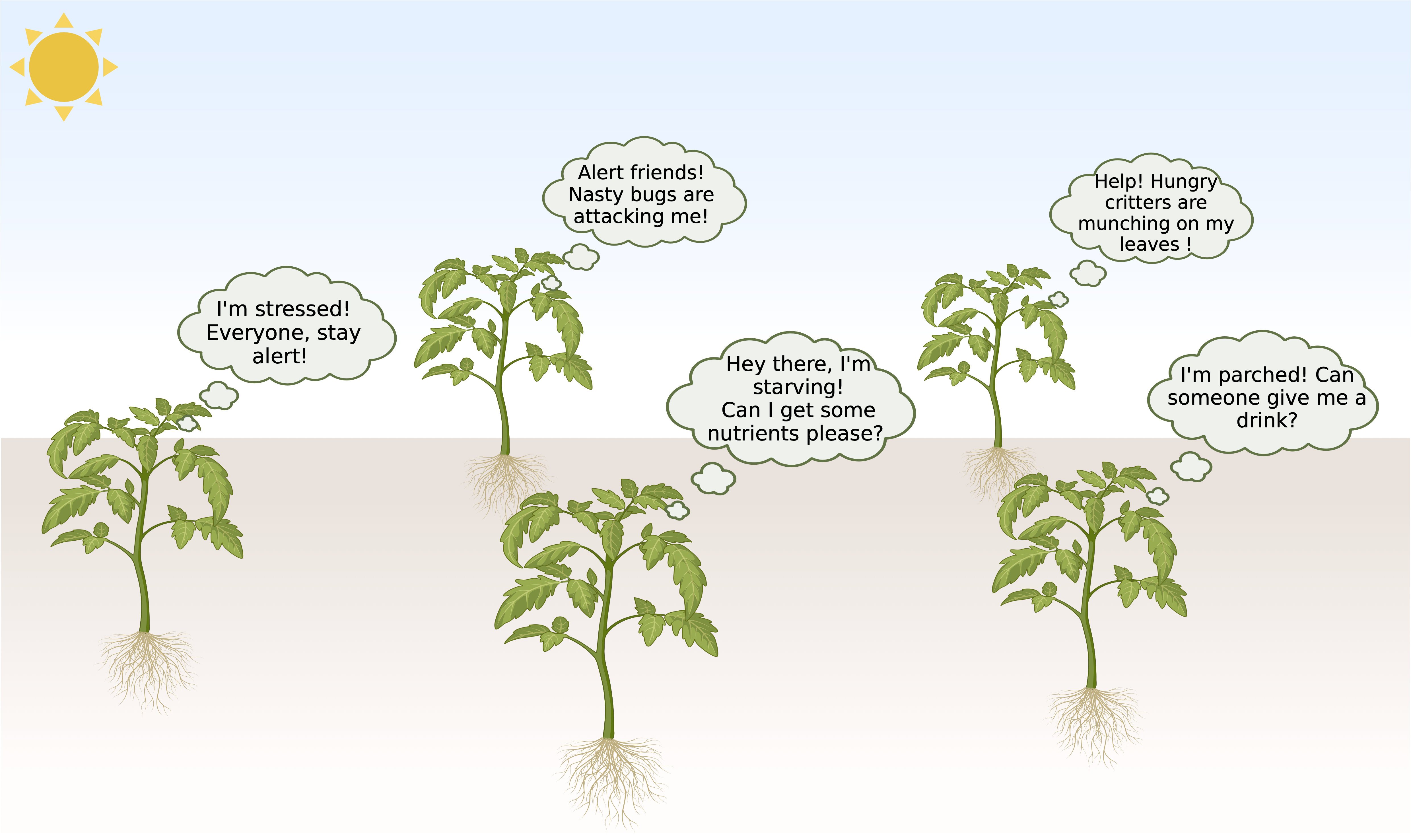}
    \caption{An illustration of interplant communication in nature, created with BioRender.com.}
    \label{fig:plantCommunication}
\end{figure}

The study of plants encompasses a broad range of topics, including diversity, photosynthesis, structure, growth and development, reproduction, learning, memory, and behavior. One significant aspect of plant behavior is their ability to communicate within their species and with other biological kingdoms, such as fungi, bacteria, and animals. This communication facilitates the enhancement of defense mechanisms against environmental stresses, adapts to varying conditions, and supports nutrient exchange and acquisition by forming network of plants \cite{babar2024sustainable, sharifi2021social, meents2020plant, gagliano2012out}. Fig. \ref{fig:plantCommunication} illustrates the interplant communication in nature.

Depending on the type of communication, plants can function either as natural receivers or transmitters. This implies that they are equipped with specialized perception mechanisms that enable them to receive a variety of stimuli. Additionally, they possess signal generation mechanisms that allow for the transmission of signals in diverse forms. The investigation of communication systems involving plants, either as transmitters or receivers, from an engineering perspective holds significant potential for various applications. Understanding and manipulating these communication channels could revolutionize our ability to interact with and interpret plant signals.

This article will explore potential applications of establishing communication links with plants. Before addressing these applications, it is essential to outline the existing communication modalities observed in nature among plants. This will lay a solid foundation for the vision we are developing. 

Electrical signaling has emerged as a prevalent mode of communication in plants, occurring both above and below ground. Researchers in \cite{volkov2019electrical}, demonstrated that plants can transmit electrical signals to one another through soil when interconnected by metallic conductors, utilizing Ag/AgCl and platinum electrodes along with electrostimulation techniques to induce and measure these signals. In another study \cite{volkov2018electrical}, they investigated the propagation of electrical signals within and between tomato plants and its functionality. This research also emphasized the plants' ability to distinguish between signals varying in shape, amplitude, and frequency. Additionally, in \cite{szechynska2022aboveground}, it has been explored that above-ground electrical communication exists among plants whose leaves were either in contact or connected by conductive materials. They employed methods such as electrodes and chlorophyll fluorescence imaging to monitor signal propagation, concluding that such signals can significantly alter photosynthesis, gene expression, and other physiological processes in both transmitting and receiving plants. 

The utilization of acoustic signals in plants for perceiving environmental cues or expressing stress responses is also increasingly recognized. Within the context of environmental perception, studies in \cite{ghosh2016corrigendum, jung2020sound}, analyze the impact of sound vibrations on plants by examining changes in gene expression, protein profiles, and hormone levels. Additionally, research in \cite{kollasch2020leaf} explores how plants respond to vibrations from various herbivorous insects, concluding that leaf vibrations consistently induce defensive responses in plants. In terms of stress expression, research in \cite{khait2023sounds} investigates how stressed plants emit detectable airborne sounds that can indicate their health and stress levels. The study reveals that these acoustic signals are informative and correlate with different stress conditions, by recording acoustic signals and applying machine learning algorithms for analysis.

Volatile Organic Compound (VOC) signaling is another prevalent method of communication among plants. Numerous studies have investigated how plants utilize VOCs for communication \cite{ninkovic2021plant, kessler2023volatile}. The body of research indicates that VOCs can significantly enhance plant resilience against biotic and abiotic stresses. This enhancement occurs through the emission of VOCs in response to stress and by priming plants for enhanced defense mechanisms \cite{jin2023volatile, midzij2022stress, aratani2023green}.

Recognizing these communication modalities as evidence, there is significant potential for intervening in the communication channels of plants and developing new applications. These applications could be built upon the concept of agent plants that we propose in this paper.

Rest of this paper is structured as follows. In Section II, we propose the concept of agent plants, which are capable of communicating with plants through every possible modality of plant communication discussed. In Section III, we present a novel idea for harvesting energy from plants using the agent plant-plant connection. In Section IV, we draw conclusions and highlight future research directions.

\section{Agent Plants}

An agent plant can be defined as an artificially created, plant-like structure that can be embedded within plant fields, such that its primary functions include communicating with surrounding plants, convey the gathered information to a central agent plant, and influence the behavior of surrounding plants to achieve specific objectives based on the information received from central agent plant.

An agent plant is expected to be able to
\begin{itemize}
    \item communicate with all plants within its identified area and monitor their status,
    \item communicate with a central agent plant to send the data of plants in its identified area,
    \item construct agent plant-plant highways to deliver chemical and electrical signals,
    \item store nutrients, drugs, and infochemicals to deliver to connected plants as needed through the highways,
    \item imitate mycorrhizal fungi-plant symbiosis for energy harvesting.
\end{itemize}

These capabilities imply that the agent plant would require a complex architecture comprising various types of components. These components can be sorted as processor unit, communication unit, power unit, storage unit. The subsequent sections explore each unit in detail. 

\subsection{Processor Unit}

A processor unit is essential for managing the functionalities of the agent plant. This unit provides instructions to other units to perform their operations and regulates the traffic of signals within the agent plant, ensuring connectivity and harmonious operation among all units. Additionally, the processor unit processes and stores information obtained from the agent plant's communication with the surrounding plants and central agent plant.

The processor unit stores information about all plants within its identified area by assigning a unique identifier to each plant and recording data such as health status, heat stress status, and nutrient needs. As the agent plant receives updates through communication channels, it records changes in each plant's status and takes necessary actions when required. Moreover, it regularly informs the central agent plant about its status and the status of the plants within its identified area.

\subsection{Communication Unit}

As discussed in the introduction, plants communicate with each other through various modalities, including electrical, acoustic, and molecular communication. Of these modalities, electrical communication occurs through the soil, as soil acts as a conductor. In contrast, acoustic communication is observed to occur through airborne channels, while molecular communication also primarily takes place through airborne channels. Therefore, communication unit of an agent plant includes transceivers for each modality : electrical transceiver, acoustic transceiver, molecular transceiver. In addition to this, to be able to establish the communication with the central agent plant, traditional electromagnetic communication techniques could be utilized with the equipped EMC transceiver.

\subsubsection{Electrical Transceiver}

An agent plant can communicate with connected plants utilizing electrical communication. For electrical signaling, a direct connection between an agent plant and each plant in its identified area is necessary. This provides more reliable and faster connection. Moreover, this connection allows the agent plant to decode information received from each plant and to encode and transmit new information whenever necessary.

\subsubsection{Acoustic Transceiver}

Acoustic signaling differs from electrical signaling among plants due to its wireless nature. An acoustic transceiver is a crucial component of the communication unit, capable of detecting sound signals released by plants into the environment and responding accordingly. Since there is no direct connection, the source of the signal must be detected. Based on the information extracted from these signals, the agent plant should take necessary precautions to ensure the secure and healthy life of the plants it monitors.

\subsubsection{Molecular Transceiver}

Molecular communication among plants is realized through Volatile Organic Compounds (VOCs). These molecules are generally released from the leaves of plants and come in different types, serving as indicators of various types of information. Detecting these molecules is a task of the agent plant by using molecular transceivers. These transceivers are also equipped with a VOC release mechanism, enabling the agent plant to transmit information signals to plants using VOCs when necessary.

\begin{figure}[t]
    \centering
    \includegraphics[scale = 0.2]{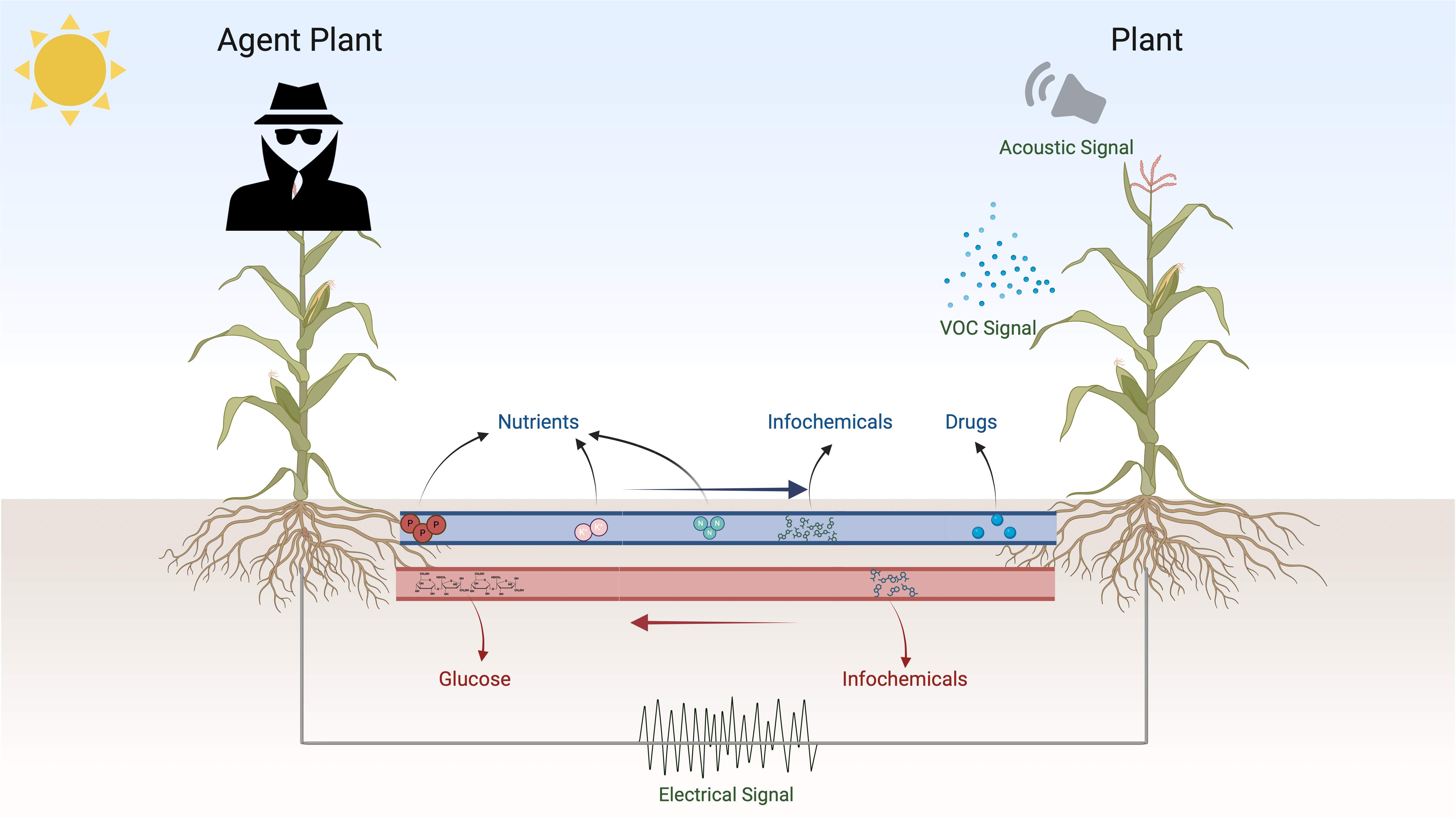}
    \caption{Agent plant - plant connection, created with BioRender.com.}
    \label{fig:agentPlantPlantCommunication}
\end{figure}

\subsection{Storage Unit}

An agent plant constantly monitors the status of plants that it has connection. In case of an alert situation, it directly transfers the necessary molecules, including nutrients, drugs, water, or infochemicals depending on the alert. These molecules are kept at the storage units of agent plants. The transfer of molecules necessitate an underground connection between the roots of the agent plant and plants to facilitate the transfer of these molecules. This underground connection is also crucial for the energy harvesting method we will discuss.

\subsection{Power Unit}

Energy is an essential component for each entity to perform its tasks. The power unit of an agent plant supports it with stored energy. For sustainability, an agent plant is expected to generate its own energy. Traditional methods, such as energy extraction from solar panels, can be employed. However, we propose an innovative method for agent plants that harvest energy by mimicking the mycorrhizal fungi-plant symbiosis. This method will be elaborated upon in the upcoming section.

Fig. \ref{fig:agentPlantPlantCommunication} illustrates the connection between an agent plant and a single plant. An agent plant can connect to multiple plants within its identified area in the same manner.

\section{Mycorrhizal Fungi and Plant Symbiosis for Energy Harvesting}

In this section, we investigate and provide a brief overview of the mycorrhizal fungi-plant symbiosis. Next, we explore oxidative respiration occurring in mitochondria as an energy harvesting technique.

\subsection{Mycorrhizal Fungi and Plant Symbiosis}

There are various types of fungi in nature, one of which is mycorrhizal fungi. The mycorrhizal fungi-plant relationship is a mutualistic symbiosis that forms in the roots of plants, where both parties benefit from the exchange of products resulting from their interaction \cite{tedersoo2020how}. The relationship begins with the germination of fungal spores, followed by the formation of germ tube hyphae that search for a host. When host plant roots are detected through chemical signals in root exudates, the fungal hyphae undergo intense branching. Upon physical contact, the fungi form specialized structures like branches. This allows further interactions from which both parties benefit \cite{seddas2009communication}.

In this symbiosis, mycorrhizal fungi significantly enhance plant growth and health by facilitating the absorption of essential nutrients such as phosphorus and nitrogen, which are often limited in soil. The ability of mycorrhizal fungi to access water and nutrients from distant soil areas increases plant resistance to various abiotic stresses, such as drought and salinity. Moreover, mycorrhizal fungi provide a protective barrier against soil-borne pathogens, enhancing the plants' defense mechanisms \cite{tedersoo2020how, simard2012mycorrhizal}. 

In return, mycorrhizal fungi benefit from their symbiotic relationship with plants by gaining access to carbohydrates produced through photosynthesis, which are essential for their growth and development. The plant roots also offer a stable and protected environment for the fungi to colonize and thrive, ensuring a reliable habitat \cite{tedersoo2020how, simard2012mycorrhizal}.

Mycorrhizal fungi form Common Mycorrhizal Networks (CMNs) by connecting the root systems of multiple plants. These networks link plants of the same or different species, enabling them to share resources and signals \cite{figueiredo2021common, newman1988mycorrhizal}. CMNs facilitate the transfer of nutrients such as carbon, nitrogen, and phosphorus between interconnected plants, allowing those with excess nutrients to share with those in deficit. Furthermore, CMNs enable plants to communicate stress signals related to pathogen attacks or herbivory. This signaling can induce defense mechanisms in neighboring plants, thereby enhancing collective resilience and defense mechanisms \cite{newman1988mycorrhizal, gilbert2017plant, babikova2013underground}.

Signaling between plants and mycorrhizal fungi occurs through the exchange of chemical signals. Plants release strigolactones into the soil, which stimulate hyphal branching and growth in mycorrhizal fungi. In response, mycorrhizal fungi release Myc factors that initiate the symbiotic relationship. In plants, the perception of Myc factors triggers calcium oscillations, which are part of the signal transduction pathway. This pathway leads to various cellular responses that facilitate fungal entry and colonization \cite{bonfante2010mechanisms, requena2007plant}.

\subsection{Respiration Based Energy Harvesting}

The process of harnessing energy from the environment or other energy sources and turning it into electrical energy is known as energy harvesting \cite{Sudevalayam2011}. Three parts make up a typical energy harvesting system: the load, the harvesting architecture, and the energy source. The energy source describes the available ambient energy that can be harnessed. The processes in a harvesting architecture are designed to harness and transform ambient energy input into electrical energy, whereas load describes the process that uses energy and serves as a sink for the energy that has been harvested.

For energy harvesting, various sources can be utilized, including solar, wind, vibration, RF, mechanical, and electrochemical energy. Each of these sources require different types of conversion mechanisms in their harvesting architecture \cite{beeby2006energy}. Among these, bioinspired conversion mechanisms draw inspiration from their biological counterparts. These mechanisms are based on the transport of various ions through ion channels, utilizing differences in the concentration of salts, protons, or electrons \cite{li2017adv}. 

In this paper, we will focus on respiration based conversion mechanisms inspired by mitochondrial respiration. The reason for this focus is that mitochondrial respiration produces chemical energy using glucose. On the other hand, the proposed agent plant is expected to extract glucose from the plants it contacts by exchanging nutrients or water. This technique mirrors the trade observed in the symbiotic relationship between mycorrhizal fungi and plants. Moreover, it introduces an innovative energy harvesting method to the literature.

An agent plant connects to each plant within its identified area through underground channels. These channels facilitate signaling to the plants to extract glucose. The extracted glucose is transported to the agent plant for further processing and conversion into electrical energy. There exists different harvesting architectures for conversion of glucose into electrical energy. These architectures are based on the cellular respiration which is a multi-step metabolic process that converts glucose into ATP. Cellular respiration converts glucose into ATP through three main stages. First, glycolysis in the cytoplasm splits glucose into two pyruvate molecules, producing ATP and NADH. The pyruvate is then transported into the mitochondria, where it undergoes oxidation to form acetyl-CoA. Next, acetyl-CoA enters the citric acid cycle in the mitochondrial matrix, generating ATP, NADH, and FADH2. Finally, in the electron transport chain, high-energy electrons from NADH and FADH2 drive ATP production through a proton gradient, with oxygen forming water as the final electron acceptor. Overall, cellular respiration produces approximately 30-38 ATP molecules from glucose \cite{urry2021campbell}.

\begin{figure}[t]
    \centering
    \includegraphics[scale = 0.3]{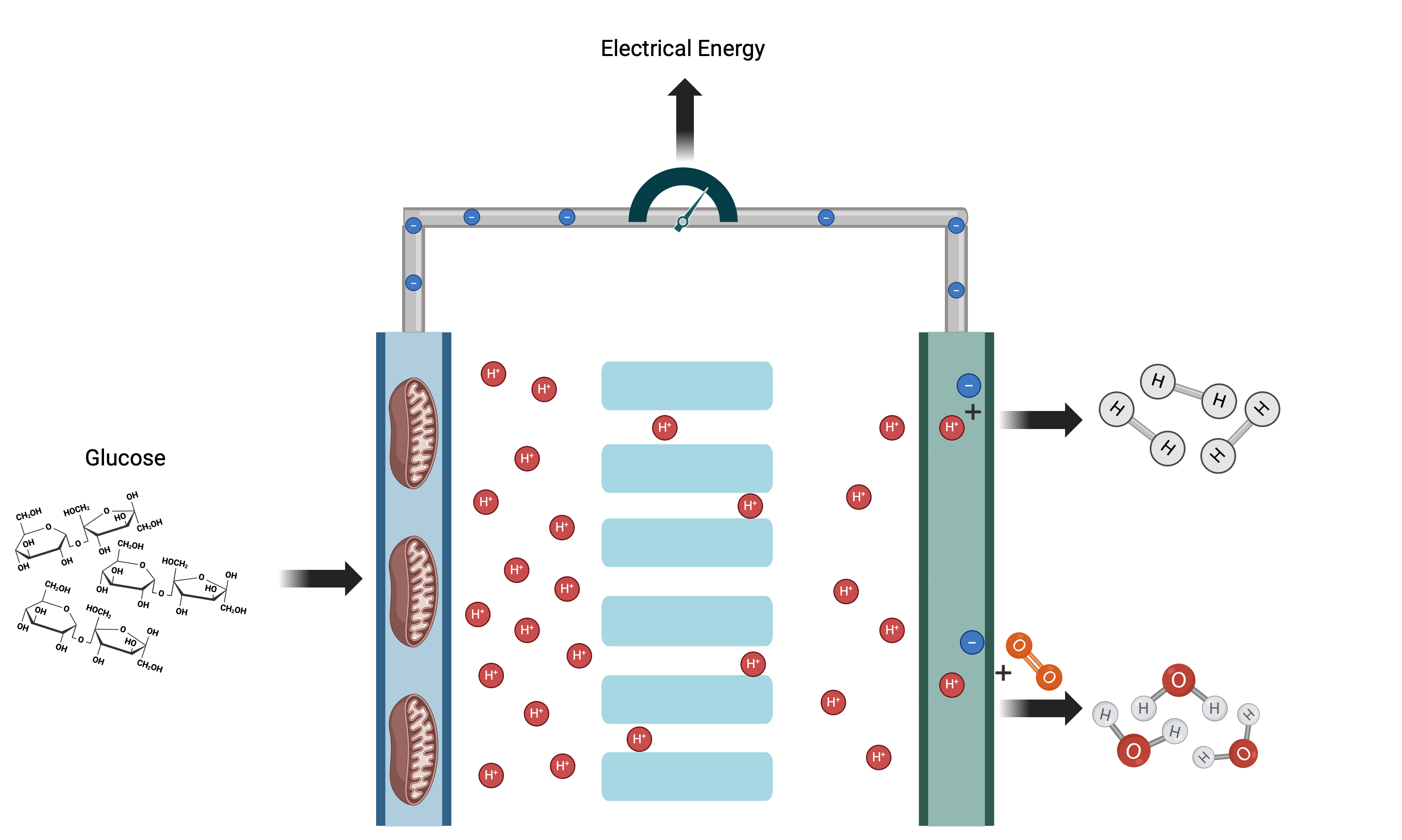}
    \caption{Harvesting architecture for respiration based biocell, created with BioRender.com.}
    \label{fig:energyConversion}
\end{figure}

Inspiring from its biological counterpart, respiration based harvesting architectures consist of three parts: an anode, a cathode, and nanochannels \cite{zhang2017adv, wang2024biological}. The anode contains mitochondria extracted from animals and is provided with glucose. The end products of interactions at the mitochondria's membrane are electrons and protons as hydrogen ions. The electrons generate current by moving to the cathode through connected wires, while the protons move to the cathode through nanochannels. These nanochannels replace the proton transport membranes found in their biological counterparts. They provide passages for protons to transport along the concentration gradient to consume electrons on the cathode. This enables a continuous and stable current (Fig. \ref{fig:energyConversion}).

In light of this architecture, a high-performance biocell was obtained in \cite{zhang2017adv} that is equipped with polyethylene terephthalate (PET) nanochannels, producing a current of approximately 3.1 mA cm\(^{-2}\), a maximum power of approximately 0.91 mW cm\(^{-2}\) at 0.35 V, and a lifetime exceeding 60 hours. By regulating the permeability area and surface charge of nanochannels, the performance of the biocell was significantly enhanced.

In another study \cite{wang2024biological}, a current density of 6.42 mA cm\(^{-2}\) and a maximum power density of 1.21 mW cm\(^{-2}\) were achieved for 8 days. This biocell used an artificial nanochannel composed of sulfonated poly(ether ether ketone) (SPEEK) fiber networks to provide rapid proton pathways. The anode was constructed with mitochondria and a cellular respiratory solution containing 8 mM malic acid, 20 mM hydroxyethyl piperazine ethyl sulfonic acid, 2 mM MgCl\(_2\), and 250 mM sucrose. 

The artificial nanochannels with properties such as ion selectivity, ion gating, and ion rectification are important components of these architectures. To further enhance the performance of these biocells, it is necessary to make smart bioinspired nanochannels more stable, reversible, and durable in extreme conditions \cite{hao2020nanochannels}. Future research on these topics can significantly enhance the power output that can be obtained from respiration based biocells.

\section{Performance Evaluation}

\subsection{Photosynthesis Based Energy Harvesting}
To evaluate the performance of respiration based biocells, we included a comparison with a more extensively studied harvesting architecture based on photosynthesis. Photosynthetic energy harvesting captures and converts the energy produced during photosynthesis into usable electrical power with high efficiency \cite{kim2021photosynthetic}. Photosynthesis involves light and dark reactions, where light energy splits water molecules, producing oxygen, protons, and electrons that synthesize ATP and NADPH, ultimately resulting in glucose production. The photosynthetic electron transport (PET) chains, which include components such as Photosystem II (PS II) and Photosystem I (PS I), play a crucial role in generating high-energy electrons. These electrons can be harnessed for electricity generation. Thus, Bio-Photovoltaic (BPV) cells leverage photosynthetic organisms or components to convert light energy into electrical energy, showcasing the potential of photosynthetic processes for efficient energy conversion.

\begin{figure}[t]
    \centering
    \includegraphics[scale = 0.18]{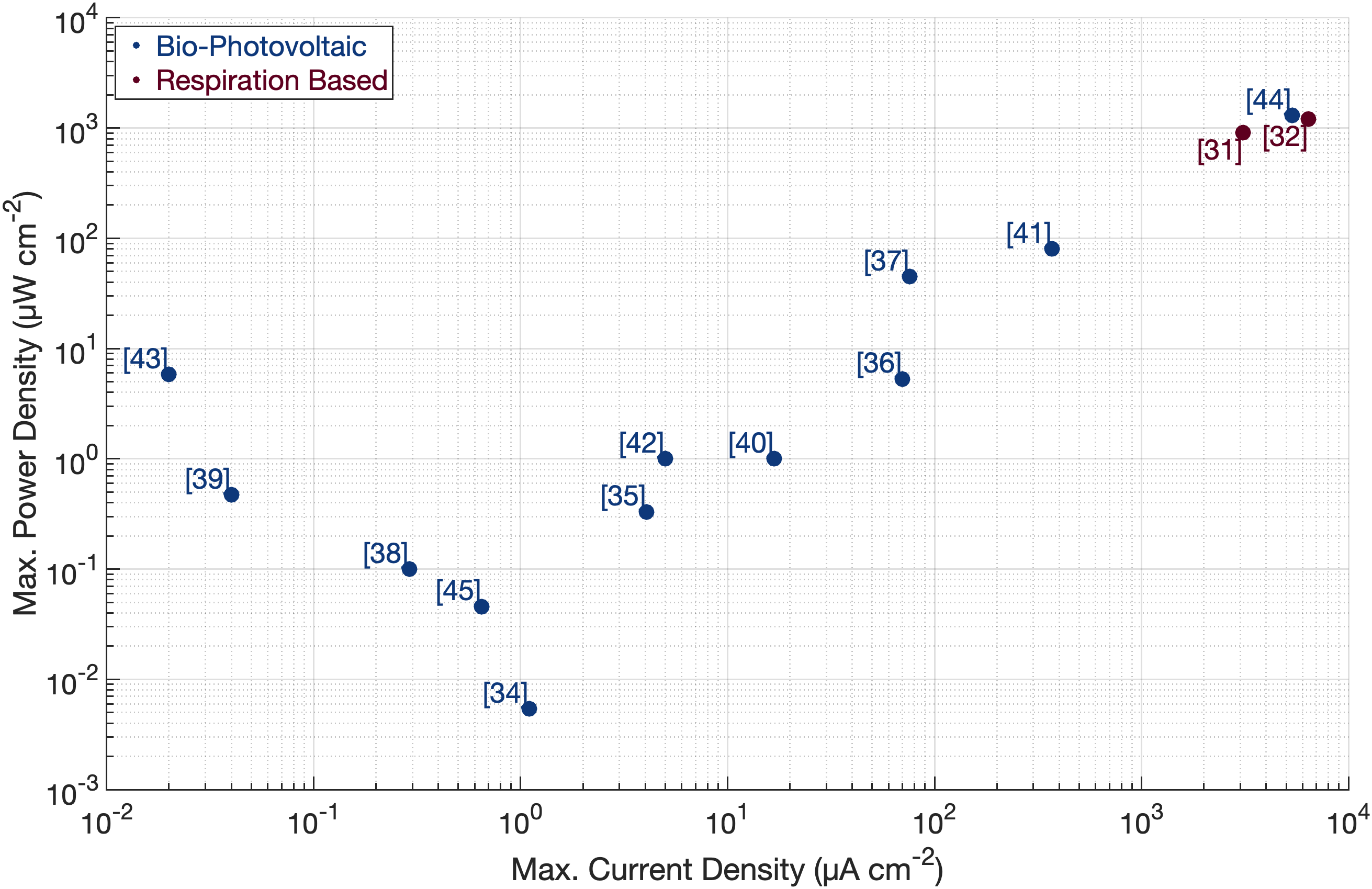}
    \caption{A comparison of Bio-Photovoltaic cells with Respiration Based cells over maximum power density and maximum current density.}
    \label{fig:comparison}
\end{figure}

Fig. \ref{fig:comparison} presents a comparison of respiration based (\cite{zhang2017adv, wang2024biological}) and photosynthetic based (\cite{33,34,35,36,37,38,39,40,41,42,43,44}) energy harvesting systems. The graph indicates that although there are fewer studies on respiration based systems, their performance outputs are superior. This implies that the approach of extracting glucose from plants, similar to how fungi do, for use in respiration based energy harvesting is promising.

On the other hand, agent plants do not need to rely on a single method of energy harvesting. They can also be equipped with photosynthetic biocells on their aboveground parts, providing supplementary energy. In this setup, while photosynthetic biocells would depend on light, respiration based systems would rely on glucose. Since glucose extraction is not dependent on light, the ability to harvest energy at night would be an additional advantage of the respiration based system. By this means, this dual approach would ensure a more consistent and reliable energy supply for an agent plant.

\subsection{Mathematical Model}

The performance of respiration based biocells primarily depends on several key parameters: the concentration of glucose molecules at the anode, nanochannel parameters such as material type, pore size, and surface charge, and the electron consumption rate at the cathode \cite{zhang2017adv}. In our study, we propose to extract glucose and transport it to the agent plant for energy harvesting. Our objective is to mathematically model the potential effect of glucose concentration on this system.

Consider an agent plant equipped with a respiration based biocell with a maximum power density of \(P_{\text{max}}\) per \(\text{cm}^2\). This agent plant is connected to \(n\) other plants. At each time interval \(T_i\), the agent plant either transports \(g_i\) moles of glucose from a plant with a probability of \(p\) or fails to transport anything with a probability of \(1 - p\). The total amount of glucose extracted at a single interval can be modeled as a binomial random variable \(G_i\) with an expected value
\begin{equation}
    E[G_i] = n \cdot g_i \cdot p.
    \label{eq:binomial}
\end{equation}
The extracted glucose molecules are then fed to the mitochondria at the anode. Since the amount of glucose fed at each time interval is a random variable, the glucose concentration can be represented as a random process \(S(t)\).

Mitochondria catabolize glucose through enzymatic reactions dependent on glucose concentration. To model the catalytic rate of mitochondrial enzymes, we adopt the Michaelis-Menten equation, as detailed in \cite{srinivasan2022guide, tseng2022kinetic, korla2013modelling}, and express the expectation of catalytic rate as
\begin{equation}
    E[V(t)] = \frac{V_{\text{max}} \cdot E[S(t)]}{K_m + E[S(t)]},
    \label{eq:michaelis-menten}
\end{equation}
where \(V_{\text{max}}\) is the maximum catalytic rate when the mitochondrial enzymes are saturated with glucose, and \(K_m\) is the Michaelis-Menten constant, representing the glucose concentration at which the reaction catalytic rate is half of \(V_{\text{max}}\). Therefore, the expected value of the glucose concentration at the anode can then be expressed as
\begin{equation}
    E[S(t)] =  \left(\sum_{i=1}^{\lfloor \frac{t}{T_i} \rfloor} E[G_i]\right) - \int_0^t E[V(\tau)] \, d\tau.
    \label{eq:est}
\end{equation}
Substituting (\ref{eq:binomial}) and (\ref{eq:michaelis-menten}) into (\ref{eq:est}), we get
\begin{equation}
    E[S(t)] = n \cdot g_i \cdot p \cdot \lfloor \frac{t}{T_i} \rfloor - \int_0^t \frac{V_{\text{max}} \cdot E[S(\tau)]}{K_m + E[S(\tau)]} \, d\tau.
\end{equation}

The ultimate goal of this model is to derive a mathematical expression for the power density of biocell as a function of time. We assume that the catalytic rate of the mitochondria is correlated with the power output. As the mitochondria operate faster, they are expected to produce more electrons and protons, thereby increasing the power output. Thus, we also state the expected power density of biocell at time \(t\) as
\begin{equation}
    E[P(t)] \propto \frac{P_{\text{max}} \cdot E[S(t)]}{K_m + E[S(t)]}.
    \label{eq:power}
\end{equation}

\begin{figure*}[htbp]
    \centering
    \begin{subfigure}[b]{0.3\textwidth}
        \centering
        \includegraphics[width=\textwidth]{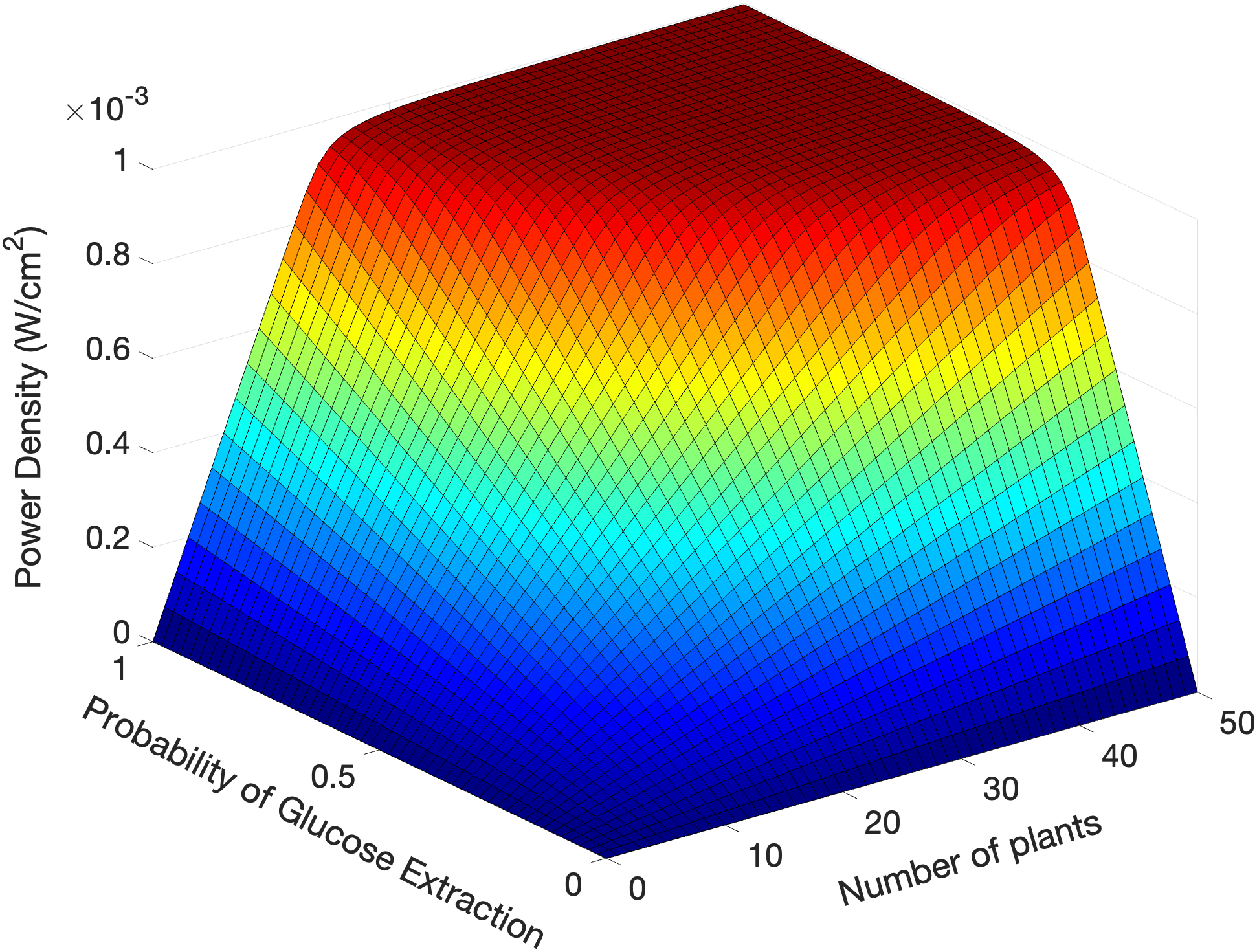}
        \caption{Effect of probability of glucose extraction, \(p\), from a single signalling event with the number of plants, \(n\), on power density at \(t = 5h, g_i = 1 \mu M, T_i = 60s\).}
    \end{subfigure}
    \hfill
    \begin{subfigure}[b]{0.3\textwidth}
        \centering
        \includegraphics[width=\textwidth]{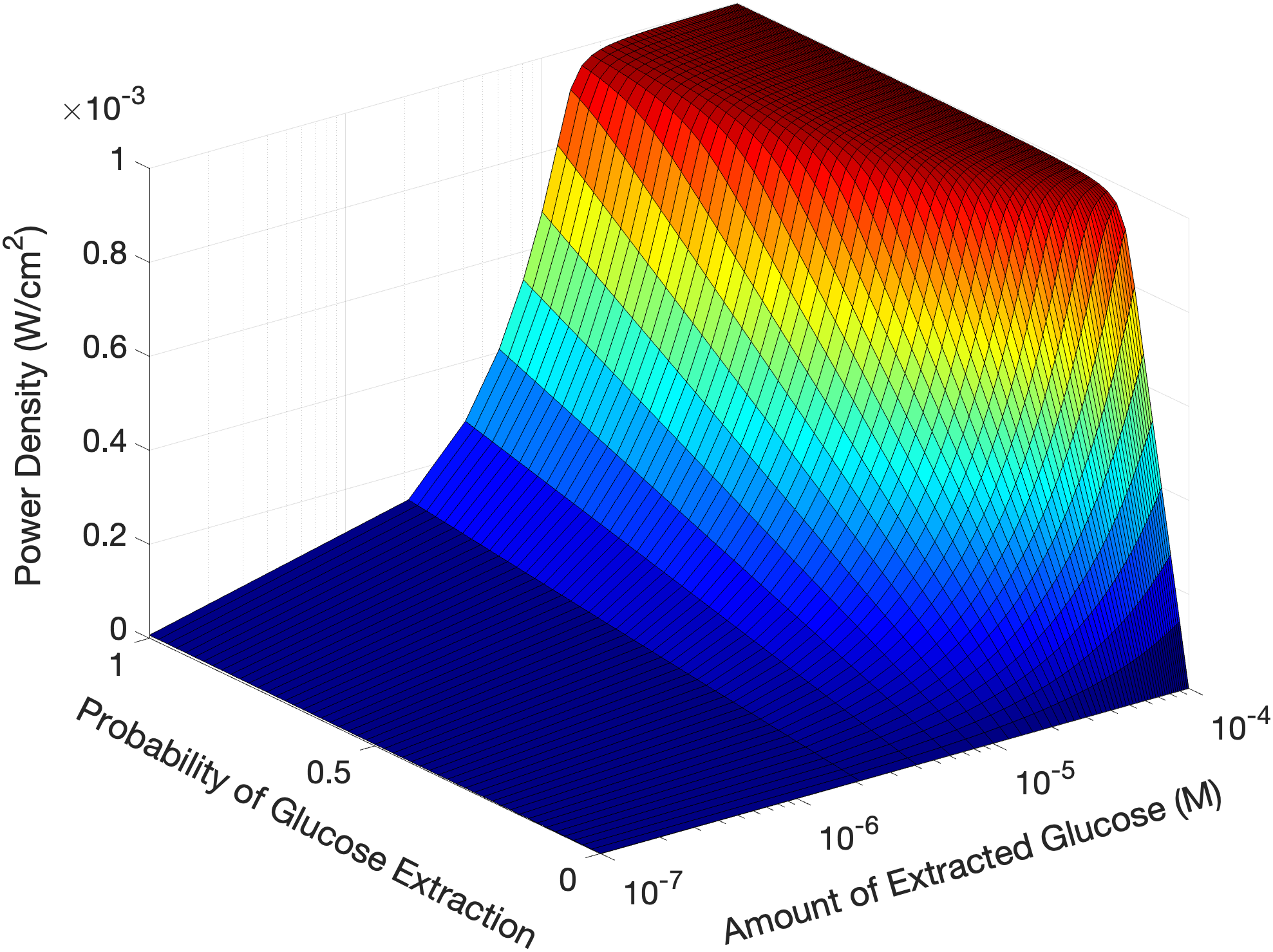}
        \caption{Effect of probability of glucose extraction, \(p\), with the amount of extracted glucose, \(g_i\), from a single signalling event, on power density at \(t = 5h, T_i = 60s, n = 1\).}
    \end{subfigure}
    \hfill
    \begin{subfigure}[b]{0.3\textwidth}
        \centering
        \includegraphics[width=\textwidth]{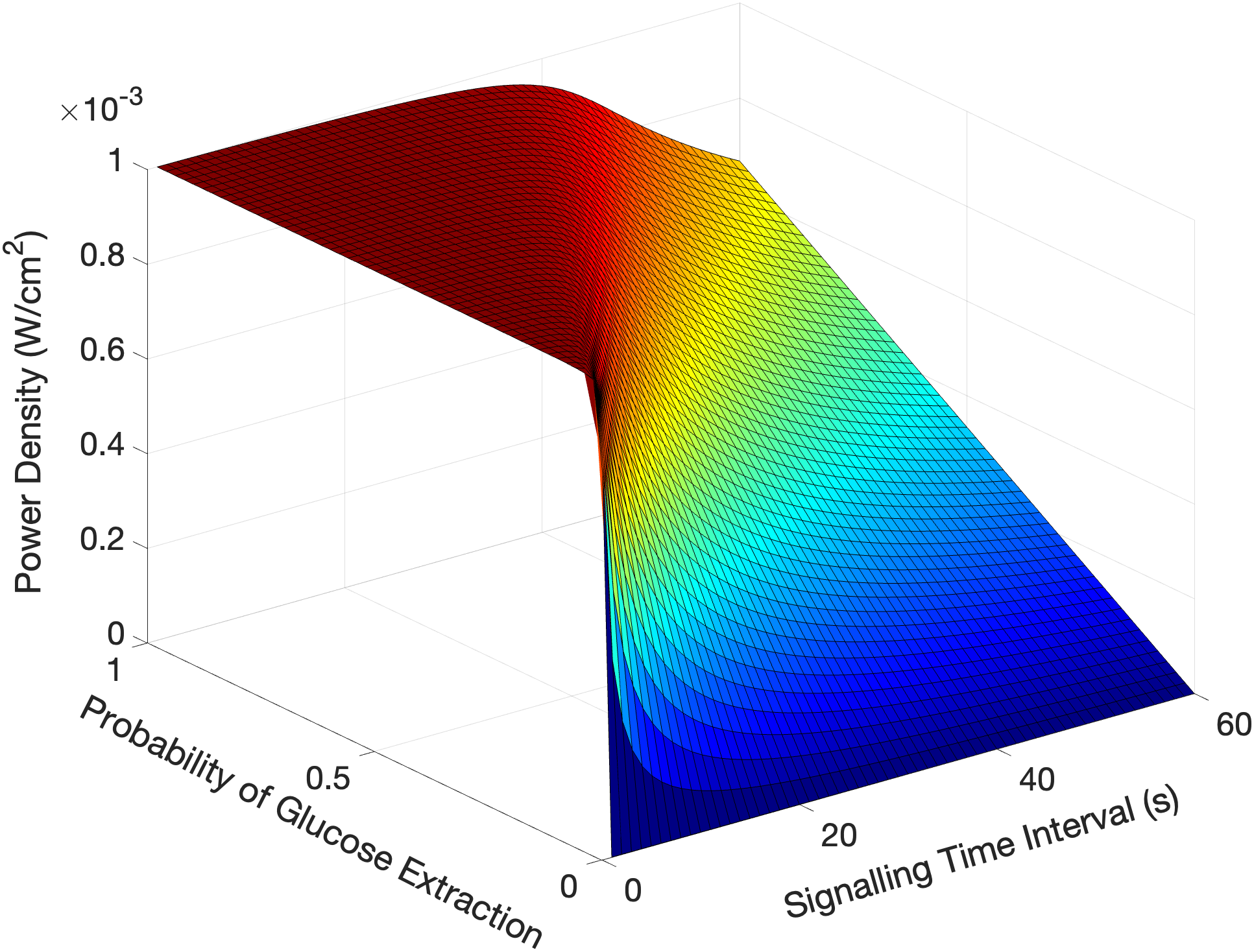}
        \caption{Effect of probability of glucose extraction, \(p\), from a single signalling event with the signalling time interval, \(T_i\), on power density at \(t = 5h, g_i = 10 \mu M, n = 1\).}
    \end{subfigure}
    \caption{Effect of system parameters on power density of the respiration based biocell.}
    \label{fig:three_figures}
\end{figure*}

\subsection{Analysis of System Parameters}

We derived an equation for the expected value of \(P(t)\). Using this equation, we analyzed the impact of various system parameters on power density of respiration based biocell. In our analyses, we focused on parameters related to the mycorrhizal fungi and plant symbiosis, as this is the novel aspect of our proposal. Therefore, we conducted analyses on the parameters \(p\), \(n\), \(g_i\), and \(T_i\). For the other parameters, we selected values that align with the experimental studies in \cite{zhang2017adv, wang2024biological, kim2021steady}: \(P_{\text{max}} = 1 \frac{mW}{cm^2}\), \(V_{max} = 0.25 \frac{\mu M}{s}\), \(K_m = 30 \mu M\).

Fig. \ref{fig:three_figures} presents the analysis results. Each analysis was conducted with two parameters, one of which is the probability of glucose extraction from a single signaling event. As observed in all the plots, an increase in the \(p\) value leads to an increase in the power output density of the biocell. This is because a higher probability facilitates more glucose transportation to the agent plant, allowing the mitochondria to operate at higher catalytic rate. This analysis highlights the importance of a reliable exchange between the agent plant and the connected plants.

In Fig. \ref{fig:three_figures}(a), the impact of the number of plants, \(n\), connected to an agent plant is analyzed. Increasing the number of connected plants enhances the total amount of glucose transported to the agent plant. This results in a higher glucose concentration at the anode and, consequently, an increase in the power density of the biocell. However, it is important to consider that increasing the number of plants also requires more energy consumption for agent plant, as the agent plant must now support more plants.

In Fig. \ref{fig:three_figures}(b), the impact of the amount of glucose extracted, \(g_i\), from a single signaling event is investigated. It is observed that the ability to operate at maximum power density increases with increasing \(g_i\), as more glucose becomes available for the agent plant. When the amount transported to the agent plant is insufficient, depleted glucose levels result in zero energy production, for instance, at values smaller than 1 \(\mu\)M for the case in the figure. According to \cite{douds2000carbon}, a plant directs 4\% to 20\% of the glucose produced by photosynthesis to mycorrhizal fungi in its roots. This underscores the importance of the amount of glucose that can be extracted from a single signaling event.

In Fig. \ref{fig:three_figures}(c), the impact of the signaling time interval, \(T_i\), has been examined. Since each signaling event represents a potential glucose extraction with a probability \(p\), the frequency of signaling becomes crucial. Therefore, an increase in \(T_i\) decreases the level of glucose concentration at the biocell, resulting in lower power density values at the output.
\section{Conclusion}

In this study, we explained the communication methods that exist between plants: electrical, acoustic, and molecular communication techniques. We then introduced agent plants, detailed their architecture. A key idea that we propose in this article was the discussion of imitating the mycorrhizal fungi and plant symbiotic relationship. We briefly mentioned the existing symbiotic relationship in nature and explored how we can utilize it for energy harvesting. 

To develop agent plants, future research needs to focus on modeling plant-to-plant communication across different modalities. Although there are existing studies on molecular communication in the plant context \cite{aktas2023odorbased}, more research is needed for all other modalities. Developing new modulation methods as in \cite{powari2024, bilgen2024odor} and signal detection techniques as in \cite{Kuscu2024, Baydas2024, koca2024modelling} are also crucial for establishing a reliable communication link with plants. On the other hand, a critical direction is realizing the idea of energy harvesting. To achieve this, extensive research on plant signaling and the production of respiration-based biocells should be conducted.

%
\begin{acks}
This work was supported in part by the AXA Research Fund (AXA Chair for Internet of Everything at Ko\c{c} University).
\end{acks}

\printbibliography

@String{Academic = "Academic Press" }

@String{Springer = "Springer-Verlag" }

@article{volkov2019electrical,
  title={Electrical signal transmission in the plant-wide web},
  author={Volkov, Alexander G. and Toole, Shannon and WaMaina, Mwangi},
  journal={Bioelectrochemistry},
  volume={129},
  year={2019},
  pages={70-78},
  issn={1567-5394},
  doi={10.1016/j.bioelechem.2019.05.003},
  url={https://doi.org/10.1016/j.bioelechem.2019.05.003}
}

@article{volkov2018electrical,
  title={Electrical signal propagation within and between tomato plants},
  author={Volkov, Alexander G. and Shtessel, Yuri B.},
  journal={Bioelectrochemistry},
  volume={124},
  year={2018},
  pages={195-205},
  issn={1567-5394},
  doi={10.1016/j.bioelechem.2018.08.001},
  url={https://doi.org/10.1016/j.bioelechem.2018.08.001}
}

@article{szechynska2022aboveground,
  title={Aboveground plant-to-plant electrical signaling mediates network acquired acclimation},
  author={Szechy{\'n}ska-Hebda, Magdalena and Lewandowska, Marta and Wito{\'n}, Dominika and Fichman, Yosef and Mittler, Ron and Karpi{\'n}ski, Stanis{\l}aw M.},
  journal={Plant Cell},
  volume={34},
  number={8},
  pages={3047-3065},
  year={2022},
  month={Jul},
  day={30},
  doi={10.1093/plcell/koac150},
  pmid={35595231},
  pmcid={PMC9338792}
}

@article{ghosh2016corrigendum,
  title={Corrigendum: Exposure to Sound Vibrations Lead to Transcriptomic, Proteomic and Hormonal Changes in Arabidopsis},
  author={Ghosh, R. and Mishra, R. C. and Choi, B. and Kwon, Y. S. and Bae, D. W. and Park, S. C. and Jeong, M. J. and Bae, H.},
  journal={Scientific Reports},
  volume={6},
  pages={37484},
  year={2016},
  month={Nov},
  day={24},
  doi={10.1038/srep37484},
  note={Erratum for: Sci Rep. 2016 Sep 26;6:33370},
  pmid={27883000},
  pmcid={PMC5122249}
}

@article{kollasch2020leaf,
  title={Leaf vibrations produced by chewing provide a consistent acoustic target for plant recognition of herbivores},
  author={Kollasch, A. M. and Abdul-Kafi, A. R. and Body, M. J. A. and Pinto, C. F. and Appel, H. M. and Cocroft, R. B.},
  journal={Oecologia},
  volume={194},
  number={1-2},
  pages={1-13},
  year={2020},
  month={Oct},
  doi={10.1007/s00442-020-04672-2},
  note={Epub 2020 Jun 12},
  pmid={32533358}
}

@article{jung2020sound,
  title={Sound Vibration-Triggered Epigenetic Modulation Induces Plant Root Immunity Against Ralstonia solanacearum},
  author={Jung, J. and Kim, S.-K. and Jung, S.-H. and Jeong, M.-J. and Ryu, C.-M.},
  journal={Frontiers in Microbiology},
  volume={11},
  year={2020},
  pages={},
  doi={10.3389/fmicb.2020.01978},
  issn={1664-302X},
  url={https://www.frontiersin.org/articles/10.3389/fmicb.2020.01978/full}
}

@article{khait2023sounds,
  title={Sounds emitted by plants under stress are airborne and informative},
  author={Khait, Ilana and Lewin-Epstein, Ori and Sharon, Rami and Saban, Keren and Goldstein, Roy and Anikster, Yosef and Zeron, Yoni and Agassy, Carmi and Nizan, Sharon and Sharabi, Guy and Perelman, Roy and Boonman, Arik and Sade, Nir and Yovel, Yossi and Hadany, Lilach},
  journal={Cell},
  volume={186},
  number={7},
  pages={1328-1336.e10},
  year={2023},
  month={Mar},
  day={30},
  doi={10.1016/j.cell.2023.03.009},
  pmid={37001499}
}

@article{aratani2023green,
  title={Green leaf volatile sensory calcium transduction in Arabidopsis},
  author={Aratani, Y. and Uemura, T. and Hagihara, T. and others},
  journal={Nature Communications},
  volume={14},
  pages={6236},
  year={2023},
  doi={10.1038/s41467-023-41589-9},
  url={https://doi.org/10.1038/s41467-023-41589-9}
}

@article{ninkovic2021plant,
  title={Plant volatiles as cues and signals in plant communication},
  author={Ninkovic, Velemir and Markovic, Dimitrije and Rensing, Michael},
  journal={Plant, Cell \& Environment},
  volume={44},
  number={4},
  pages={1030-1043},
  year={2021},
  month={Apr},
  doi={10.1111/pce.13910},
  note={Epub 2020 Oct 26},
  pmid={33047347},
  pmcid={PMC8048923}
}

@article{midzij2022stress,
  title={Stress-Induced Volatile Emissions and Signalling in Inter-Plant Communication},
  author={Midzij, J. and Jeffery, D. W. and Baumann, U. and Rogiers, S. and Tyerman, S. D. and Pagay, V.},
  journal={Plants},
  volume={11},
  pages={2566},
  year={2022},
  doi={10.3390/plants11192566},
  url={https://doi.org/10.3390/plants11192566},
  note={Published: 29 September 2022},
  editor={Academic Editors: Frantisek Baluska and Gustavo Maia Souza}
}

@article{jin2023volatile,
  title={Volatile compound-mediated plant–plant interactions under stress with the tea plant as a model},
  author={Jin, Jieyang and Zhao, Mingyue and Jing, Tingting and Zhang, Mengting and Lu, Mengqian and Yu, Guomeng and Wang, Jingming and Guo, Danyang and Pan, Yuting and Hoffmann, Timothy D and Schwab, Wilfried and Song, Chuankui},
  journal={Horticulture Research},
  volume={10},
  number={9},
  pages={uhad143},
  year={2023},
  month={Sep},
  doi={10.1093/hr/uhad143},
  url={https://doi.org/10.1093/hr/uhad143}
}

@article{kessler2023volatile,
  title={Volatile-mediated plant–plant communication and higher-level ecological dynamics},
  author={Kessler, Andr{\'e} and Mueller, Michael B. and Kalske, Aino and Chaut{\'a}, Alexander},
  journal={Current Biology},
  volume={33},
  number={11},
  pages={R519-R529},
  year={2023},
  doi={10.1016/j.cub.2023.04.025},
  issn={0960-9822},
  url={https://www.sciencedirect.com/science/article/pii/S0960982223004736}
}

@article{tedersoo2020how,
  title={How mycorrhizal associations drive plant population and community biology},
  author={Tedersoo, Leho and Bahram, Mohammad and Zobel, Martin},
  journal={Science},
  volume={367},
  number={6480},
  pages={eaba1223},
  year={2020},
  month={Feb},
  doi={10.1126/science.aba1223},
  pmid={32079744}
}

@article{simard2012mycorrhizal,
  title={Mycorrhizal networks: Mechanisms, ecology and modelling},
  author={Simard, Suzanne W. and Beiler, Kevin J. and Bingham, Marcus A. and Deslippe, Julie R. and Philip, Leanne J. and Teste, Fran{\c{c}}ois P.},
  journal={Fungal Biology Reviews},
  volume={26},
  number={1},
  pages={39-60},
  year={2012},
  issn={1749-4613},
  doi={10.1016/j.fbr.2012.01.001},
  url={https://www.sciencedirect.com/science/article/pii/S1749461312000048}
}

@incollection{seddas2009communication,
  title={Communication and Signaling in the Plant–Fungus Symbiosis: The Mycorrhiza},
  author={Seddas, P. and Gianinazzi-Pearson, V. and Schoefs, B. and K{\"u}ster, H. and Wipf, D.},
  booktitle={Plant-Environment Interactions},
  editor={Balu{\v{s}}ka, Franti{\v{s}}ek},
  series={Signaling and Communication in Plants},
  pages={79-97},
  year={2009},
  publisher={Springer},
  address={Berlin, Heidelberg},
  doi={10.1007/978-3-540-89230-4_3},
  url={https://doi.org/10.1007/978-3-540-89230-4_3}
}

@article{figueiredo2021common,
  title={Common Mycorrhizae Network: A Review of the Theories and Mechanisms Behind Underground Interactions},
  author={Figueiredo, Amanda F. and Boy, Jens and Guggenberger, Georg},
  journal={Frontiers in Fungal Biology},
  volume={2},
  pages={735299},
  year={2021},
  doi={10.3389/ffunb.2021.735299}
}

@incollection{newman1988mycorrhizal,
  title={Mycorrhizal Links Between Plants: Their Functioning and Ecological Significance},
  author={Newman, E. I.},
  booktitle={Advances in Ecological Research},
  editor={Begon, M. and Fitter, A. H. and Ford, E. D. and Macfadyen, A.},
  volume={18},
  pages={243-270},
  year={1988},
  publisher={Academic Press},
  issn={0065-2504},
  isbn={9780120139187},
  doi={10.1016/S0065-2504(08)60182-8},
  url={https://www.sciencedirect.com/science/article/pii/S0065250408601828}
}

@incollection{gilbert2017plant,
  title={Chapter Four - Plant–Plant Communication Through Common Mycorrhizal Networks},
  author={Gilbert, L. and Johnson, D.},
  booktitle={Advances in Botanical Research},
  editor={Becard, Guillaume},
  volume={82},
  pages={83-97},
  year={2017},
  publisher={Academic Press},
  issn={0065-2296},
  isbn={9780128014318},
  doi={10.1016/bs.abr.2016.09.001},
  url={https://www.sciencedirect.com/science/article/pii/S0065229616300878}
}

@article{babikova2013underground,
  title={Underground signals carried through common mycelial networks warn neighbouring plants of aphid attack},
  author={Babikova, Zdenka and Gilbert, Lucy and Bruce, Toby J. A. and Birkett, Michael and Caulfield, John C. and Woodcock, Christine and Pickett, John A. and Johnson, David},
  journal={Ecology Letters},
  volume={16},
  number={7},
  pages={835-843},
  year={2013},
  doi={10.1111/ele.12115},
  url={https://doi.org/10.1111/ele.12115},
  eprint={https://onlinelibrary.wiley.com/doi/pdf/10.1111/ele.12115}
}

@article{bonfante2010mechanisms,
  title={Mechanisms underlying beneficial plant–fungus interactions in mycorrhizal symbiosis},
  author={Bonfante, P. and Genre, A.},
  journal={Nature Communications},
  volume={1},
  pages={48},
  year={2010},
  doi={10.1038/ncomms1046},
  url={https://doi.org/10.1038/ncomms1046}
}

@article{requena2007plant,
  title={Plant signals and fungal perception during arbuscular mycorrhiza establishment},
  author={Requena, N. and Serrano, E. and Ocón, A. and Breuninger, M.},
  journal={Phytochemistry},
  volume={68},
  number={1},
  pages={33-40},
  year={2007},
  month={Jan},
  doi={10.1016/j.phytochem.2006.09.036},
  note={Epub 2006 Nov 13},
  pmid={17095025}
}

@article{gagliano2012out,
  title={Out of Sight but Not out of Mind: Alternative Means of Communication in Plants},
  author={Gagliano, M. and Renton, M. and Duvdevani, N. and Timmins, M. and Mancuso, S.},
  journal={PLoS ONE},
  volume={7},
  number={5},
  pages={e37382},
  year={2012},
  doi={10.1371/journal.pone.0037382},
  url={https://doi.org/10.1371/journal.pone.0037382}
}

@article{meents2020plant,
  title={Plant–Plant Communication: Is There a Role for Volatile Damage-Associated Molecular Patterns?},
  author={Meents, Angela K. and Mithöfer, Axel},
  journal={Frontiers in Plant Science},
  volume={11},
  pages={583275},
  year={2020},
  doi={10.3389/fpls.2020.583275},
  url={https://doi.org/10.3389/fpls.2020.583275}
}

@article{sharifi2021social,
  title={Social networking in crop plants: Wired and wireless cross-plant communications},
  author={Sharifi, Rouhallah and Ryu, Choong-Min},
  journal={Plant, Cell \& Environment},
  volume={44},
  pages={1095-1110},
  year={2021},
  doi={10.1111/pce.13966},
  url={https://doi.org/10.1111/pce.13966}
}

@misc{babar2024sustainable,
      title={Sustainable and Precision Agriculture with the Internet of Everything (IoE)}, 
      author={Adil Z Babar and Ozgur B. Akan},
      year={2024},
      eprint={2404.06341},
      archivePrefix={arXiv},
      primaryClass={eess.SP}
}

@ARTICLE{Sudevalayam2011,
  author={Sudevalayam, Sujesha and Kulkarni, Purushottam},
  journal={IEEE Communications Surveys \& Tutorials}, 
  title={Energy Harvesting Sensor Nodes: Survey and Implications}, 
  year={2011},
  volume={13},
  number={3},
  pages={443-461},
  doi={10.1109/SURV.2011.060710.00094}}

@article{beeby2006energy,
  title={Energy harvesting vibration sources for microsystems applications},
  author={Beeby, S P and Tudor, M J and White, N M},
  journal={Measurement Science and Technology},
  volume={17},
  number={12},
  pages={R175},
  year={2006},
  month={Oct},
  doi={10.1088/0957-0233/17/12/R01},
  url={https://dx.doi.org/10.1088/0957-0233/17/12/R01}
}

@article{zhang2017adv,
  title={AHigh-Performance Respiration-Based Biocell Using Artificial Nanochannel Regulation},
  author={Zhang, Q. and Li, X. and Chen, Y. and Zhang, Q. and Liu, H. and Zhai, J. and Yang, X.},
  journal={Advanced Materials},
  volume={29},
  pages={1606871},
  year={2017},
  doi={10.1002/adma.201606871},
  url={https://doi.org/10.1002/adma.201606871}
}

@article{li2017adv,
  title={Smart Bioinspired Nanochannels and their Applications in Energy-Conversion Systems},
  author={Li, R. and Fan, X. and Liu, Z. and Zhai, J.},
  journal={Advanced Materials},
  volume={29},
  pages={1702983},
  year={2017},
  doi={10.1002/adma.201702983},
  url={https://doi.org/10.1002/adma.201702983}
}

@article{wang2024biological,
  title={Biological electricity generation system based on mitochondria-nanochannel-red blood cells},
  author={Wang, Yuting and Chen, Huaxiang and Yang, Xiaoda and Diao, Xungang and Zhai, Jin},
  journal={Nanoscale},
  volume={16},
  number={15},
  pages={7559-7565},
  year={2024},
  doi={10.1039/D3NR05879D},
  publisher={The Royal Society of Chemistry},
  url={http://dx.doi.org/10.1039/D3NR05879D}
}

@incollection{urry2021campbell,
  title={Cellular Respiration and Fermentation},
  author={Urry, Lisa A. and Cain, Michael L. and Wasserman, Steven A. and Minorsky, Peter V. and Orr, Rebecca},
  booktitle={Campbell Biology},
  edition={12th},
  year={2021},
  publisher={Pearson},
  address={New York},
  isbn={9780135188743},
  chapter={9}
}

@article{hao2020nanochannels,
  title={Nanochannels regulating ionic transport for boosting electrochemical energy storage and conversion: a review},
  author={Hao, Zhendong and Zhang, Qianqian and Xu, Xiaolong and Zhao, Qing and Wu, Congrong and Liu, Jingbing and Wang, Hao},
  journal={Nanoscale},
  volume={12},
  number={30},
  pages={15923--15943},
  year={2020},
  doi={10.1039/D0NR02464C},
  publisher={The Royal Society of Chemistry},
  url={http://dx.doi.org/10.1039/D0NR02464C}
}

@article{kim2021photosynthetic,
  title={Photosynthetic Nanomaterial Hybrids for Bioelectricity and Renewable Energy Systems},
  author={Kim, Y. J. and Hong, H. and Yun, J. H. and Kim, S. I. and Jung, H. Y. and Ryu, W.},
  journal={Advanced Materials},
  volume={33},
  pages={2005919},
  year={2021},
  doi={10.1002/adma.202005919},
  url={https://doi.org/10.1002/adma.202005919}
}

@article{33,
  title={A MEMS photosynthetic electrochemical cell powered by subcellular plant photosystems},
  author={Lam, K. B. and Johnson, E. A. and Chiao, M. and Lin, L.},
  journal={Journal of Microelectromechanical Systems},
  volume={15},
  pages={1243},
  year={2006}
}

@article{34,
  title={Conjugated Polymer Enhanced Photoelectric Response of Self-Circulating Photosynthetic Bioelectrochemical Cell},
  author={Zhou, X. and Gai, P. and Zhang, P. and Sun, H. and Lv, F. and Liu, L. and Wang, S.},
  journal={ACS Applied Materials \& Interfaces},
  volume={11},
  pages={38993},
  year={2019}
}

@article{35,
  title={High photo-electrochemical activity of thylakoid–carbon nanotube composites for photosynthetic energy conversion},
  author={Calkins, J. O. and Umasankar, Y. and O’Neill, H. and Ramasamy, R. P.},
  journal={Energy \& Environmental Science},
  volume={6},
  pages={1891},
  year={2013}
}

@article{36,
  title={A Z‐Scheme‐Inspired Photobioelectrochemical H2O/O2 Cell with a 1 V Open‐Circuit Voltage Combining Photosystem II and PbS Quantum Dots},
  author={Riedel, M. and Wersig, J. and Ruff, A. and Schuhmann, W. and Zouni, A. and Lisdat, F.},
  journal={Angewandte Chemie},
  volume={131},
  pages={811},
  year={2019}
}

@article{37,
  title={Solid-state biophotovoltaic cells containing photosystem I},
  author={Gordiichuk, P. I. and Wetzelaer, G. J. A. and Rimmerman, D. and Gruszka, A. and de Vries, J. W. and Saller, M. and Gautier, D. A. and Catarci, S. and Pesce, D. and Richter, S.},
  journal={Advanced Materials},
  volume={26},
  pages={4863},
  year={2014}
}

@article{38,
  title={Surface morphology and surface energy of anode materials influence power outputs in a multi-channel mediatorless bio-photovoltaic (BPV) system},
  author={Bombelli, P. and Zarrouati, M. and Thorne, R. J. and Schneider, K. and Rowden, S. J. and Ali, A. and Yunus, K. and Cameron, P. J. and Fisher, A. C. and Wilson, D. I.},
  journal={Physical Chemistry Chemical Physics},
  volume={14},
  pages={12221},
  year={2012}
}

@article{39,
  title={Characteristics of the photosynthesis microbial fuel cell with a Spirulina platensis biofilm},
  author={Lin, C.-C. and Wei, C.-H. and Chen, C.-I. and Shieh, C.-J. and Liu, Y.-C.},
  journal={Bioresource Technology},
  volume={135},
  pages={640},
  year={2013}
}

@article{40,
  title={Anomalous power enhancement of biophotovoltaic cell},
  author={Kim, M. J. and Bai, S. J. and Youn, J. R. and Song, Y. S.},
  journal={Journal of Power Sources},
  volume={412},
  pages={301},
  year={2019}
}

@article{41,
  title={Photosynthetic biofilms in pure culture harness solar energy in a mediatorless bio-photovoltaic cell (BPV) system},
  author={McCormick, A. J. and Bombelli, P. and Scott, A. M. and Philips, A. J. and Smith, A. G. and Fisher, A. C. and Howe, C. J.},
  journal={Energy \& Environmental Science},
  volume={4},
  pages={4699},
  year={2011}
}

@article{42,
  title={A photosynthetic-plasmonic-voltaic cell: Excitation of photosynthetic bacteria and current collection through a plasmonic substrate},
  author={Samsonoff, N. and Ooms, M. D. and Sinton, D.},
  journal={Applied Physics Letters},
  volume={104},
  pages={043704},
  year={2014}
}

@article{43,
  title={Bilayer Chlorophyll-Based Biosolar Cells Inspired from the Z-Scheme Process of Oxygenic Photosynthesis},
  author={Duan, S. and Dall’Agnese, C. and Chen, G. and Wang, X.-F. and Tamiaki, H. and Yamamoto, Y. and Ikeuchi, T. and Sasaki, S.-i.},
  journal={ACS Energy Letters},
  volume={3},
  pages={1708},
  year={2018}
}

@article{44,
  title={Effect of different irradiance levels on bioelectricity generation from algal biophotovoltaic (BPV) devices},
  author={Thong, C. H. and Phang, S. M. and Ng, F. L. and Periasamy, V. and Ling, T. C. and Yunus, K. and Fisher, A. C.},
  journal={Energy Science \& Engineering},
  volume={7},
  pages={2086},
  year={2019}
}

@article{srinivasan2022guide,
  title={A guide to the Michaelis–Menten equation: steady state and beyond},
  author={Srinivasan, B.},
  journal={FEBS Journal},
  volume={289},
  pages={6086-6098},
  year={2022},
  doi={10.1111/febs.16124},
  url={https://doi.org/10.1111/febs.16124}
}

@article{tseng2022kinetic,
  title={Kinetic Mathematical Modeling of Oxidative Phosphorylation in Cardiomyocyte Mitochondria},
  author={Tseng, Wen-Wei and Wei, An-Chi},
  journal={Cells},
  volume={11},
  pages={4020},
  year={2022},
  doi={10.3390/cells11244020},
  url={https://doi.org/10.3390/cells11244020}
}

@article{korla2013modelling,
  title={Modelling the Krebs cycle and oxidative phosphorylation},
  author={Korla, Kalyani and Mitra, Chanchal K.},
  journal={Journal of Biomolecular Structure and Dynamics},
  year={2013},
  doi={10.1080/07391102.2012.762723}
}

@article{kim2021steady,
  title={Steady-state kinetic analysis of mitochondrial respiratory enzymes from bovine heart mitochondria},
  author={Kim, D. and Ko, E. and Choi, M. and others},
  journal={Applied Biological Chemistry},
  volume={64},
  pages={54},
  year={2021},
  doi={10.1186/s13765-021-00626-1},
  url={https://doi.org/10.1186/s13765-021-00626-1}
}

@incollection{douds2000carbon,
  title={Carbon partitioning, cost and metabolism of arbuscular mycorrhizae in arbuscular mycorrhizas: physiology and function},
  author={Douds, D.D. and Pfeffer, P.E. and Shachar-Hill, Y.},
  booktitle={Arbuscular Mycorrhizas: Molecular Biology and Physiology},
  editor={Kapulnick, Y. and Douds, D.D. Jr.},
  year={2000},
  publisher={Kluwer Academic Publishers},
  address={Dordrecht, The Netherlands},
  note={in press}
}

@article{Kuscu2024,
  author={Civas, Meltem and Kuscu, Murat and Akan, Ozgur B.},
  journal={IEEE Transactions on Communications}, 
  title={Frequency-Domain Detection for Molecular Communication with Cross-Reactive Receptors}, 
  year={2024},
  volume={},
  number={},
  pages={1-1},
  doi={10.1109/TCOMM.2024.3381703}
}

@ARTICLE{Baydas2024,
  author={Baydas, O. Tansel and Akan, Ozgur B.},
  journal={IEEE Transactions on Molecular, Biological, and Multi-Scale Communications}, 
  title={Received Signal and Channel Parameter Estimation in Molecular Communications}, 
  year={2024},
  volume={10},
  number={1},
  pages={92-97},
  doi={10.1109/TMBMC.2023.3342731}
}

@misc{bilgen2024odor,
      title={Odor Perceptual Shift Keying (OPSK) for Odor-Based Molecular Communication}, 
      author={Fatih E. Bilgen and Ahmet B. Kilic and Ozgur B. Akan},
      year={2024},
      eprint={2402.11346},
      archivePrefix={arXiv},
}

@misc{koca2024modelling,
      title={Modelling 1D Partially Absorbing Boundaries for Brownian Molecular Communication Channels}, 
      author={Caglar Koca and Ozgur B. Akan},
      year={2024},
      eprint={2402.15888},
      archivePrefix={arXiv},
}

@misc{aktas2023odorbased,
      title={Odor-Based Molecular Communications: State-of-the-Art, Vision, Challenges, and Frontier Directions}, 
      author={Dilara Aktas and Beyza Ezgi Ortlek and Meltem Civas and Elham Baradari and Ayse Sila Okcu and Melanie Whitfield and Oktay Cetinkaya and Ozgur Baris Akan},
      year={2023},
      eprint={2311.17727},
      archivePrefix={arXiv},
}

@ARTICLE{powari2024,
  author={Powari, Aditya and Akan, Ozgur B.},
  journal={IEEE Transactions on Molecular, Biological, and Multi-Scale Communications}, 
  title={Odor Intensity Shift Keying (OISK) and Channel Capacity of Odor-Based Molecular Communications in Internet of Everything}, 
  year={2024},
  volume={},
  number={},
  pages={1-1},
  doi={10.1109/TMBMC.2024.3408063}
}
\end{document}